\documentclass[twocolumn,prl,amsmath,amssymb]{revtex4}
\usepackage{dcolumn}
\usepackage{graphicx}
\usepackage{stmaryrd}
\usepackage{bm}%

\def \ket #1{\vert #1 \rangle}

\def \brk #1#2{\langle #1\vert #2 \rangle}

\begin{document}
\title{Looking for Light Pseudoscalar Bosons in the Binary Pulsar System J0737-3039}

\author{Arnaud Dupays}
\affiliation{
Laboratoire Collisions, Agr\'egats, R\'eactivit\'e, IRSAMC,
CNRS/UPS, 31062 Toulouse, France
}
\author{Carlo Rizzo}
\affiliation{
Laboratoire Collisions, Agr\'egats, R\'eactivit\'e, IRSAMC,
CNRS/UPS, 31062 Toulouse, France
}
\author{Marco Roncadelli}
\affiliation{INFN, Sezione di Pavia, Via A. Bassi 6, I-27100 Pavia,
  Italy
}
\author{Giovanni F. Bignami}
\affiliation{
Centre d'Etude Spatiale des Rayonnements,
CNRS/UPS, 31401 Toulouse, France \\
INAF/IASF, Sezione di Milano, Italy, and Dipartimento di Fisica Nucleare e Teorica, Universit\`a di Pavia, and INFN, Sezione di Pavia, Pavia, Italy
}

\date{\today}
 
\begin{abstract}
 
We present numerical calculations of the photon-light-pseudoscalar-boson conversion in the recently discovered binary pulsar system J0737-3039. Light pseudoscalar bosons (LPBs) oscillate into photons in the presence of strong magnetic fields. In the context of this binary pulsar system, this phenomenon attenuates the light beam emitted by one of the pulsars, when the light ray goes through the magnetosphere of the companion pulsar. We show that such an effect is observable in the gamma-ray band since the binary pulsar is seen almost edge-on, depending on 
the value of the LPB mass and on the strenght of its two-photon coupling. Our results are surprising in that they show a very sharp and significant (up to 50\%) transition probability in the gamma-ray ($>$ tens of MeV) domain. The observations can be performed by the upcoming NASA GLAST mission.

\end{abstract}

\maketitle

{\it Light pseudoscalar bosons} (LPBs) coupled to two photons have attracted considerable interest in the last few years and their implications for elementary-particle phenomenology, astrophysics and cosmology have been thoroughly investigated~\cite{Raffelt 1996,graffelt,cast}. Specifically, the LPB mass $m$ is assumed to be $m < 1 \, {\rm eV}$ and the LPB-photon-photon coupling is described by the interaction lagrangian
\begin{equation}
\label{eq1}
{\cal L}_{int} = - \frac{1}{4 M} \, F^{\mu \nu} \, \Tilde{F}_{\mu \nu} \, \phi = \frac{1}{M} \, 
{\bf E} \cdot {\bf B} \, \phi~,
\end{equation}
where $\phi$ denotes the LPB field, $F^{\mu \nu}$ is the usual electromagnetic field strength 
($\Tilde{F}_{\mu \nu}$ is its dual), and $M$ is the inverse coupling constant with the dimension of a mass. Natural Lorentz-Heaviside units with $\hbar=c=1$ are employed throughout.

Because the mass eigenstates of the LPB-photon system differ from the
corresponding interaction eigenstates, interconversion takes place, a phenomenon quite similar in nature to flavour oscillations for massive neutrinos. However, since the off-diagonal term (\ref{eq1}) involves ``two photons'', one of them actually corresponds to an external field. So, 
photon-LPB oscillations occur only in the presence of external electric or magnetic fields. 

A particular case of LPB is the standard {\it axion},
namely the pseudo-Goldstone boson associated with the Peccei-Quinn
$U(1)$ global symmetry invented to solve the strong CP
problem~\cite{axion}. Although the exact value of the axion mass $m$
is model-dependent, generally one finds $m \simeq 0.7 \, (10^{10} \,
{\rm GeV}/M) \, {\rm eV}$. Besides providing a natural candidate for
nonbaryonic dark matter~\cite{Turner}, the standard axion can in
principle be detected with high-precision laboratory experiments
thanks to its two-photon coupling in eq. (\ref{eq1}). In this connection, two
strategies have been proposed. One is based on the resonant
axion-photon conversion inside a tunable microwave
cavity~\cite{Sikivie}, while the other relies upon the induced
ellipticity and change of polarization plane of an initially
linearly-polarized laser beam~\cite{Maiani}. In either case, a strong
external magnetic field must be present. Generic LPBs with a two-photon coupling (\ref{eq1}) 
can also be discovered in this way, provided of course that their mass $m$ and inverse couplig constant $M$ fall into suitable ranges which depend on the experimental arrangement.

Quite recently, using an experimental setup based on
ref. \cite{Maiani}, the PVLAS collaboration has observed an
unexplained effect that can be interpreted in terms of photon oscillations into a LPB with mass $m \simeq 1.0 \cdot 10^{- 3} \, {\rm eV}$ and inverse coupling constant $M \simeq 3.8
\cdot 10^{5} \, {\rm GeV}$ \cite{pvlas}. Clearly these values prevent the LPB in question to be the standard axion. 

On the other hand, LPBs are copiously produced in the core of stars, 
owing again to the two-photon coupling (\ref{eq1}). Therefore, under the assumption that
LPBs escape from stars, energy-loss arguments put strong bounds on
$M$. Typically, one gets $M > 10^{10} \, {\rm GeV}$~\cite{graffelt}. This conclusion is further supported by the result of the CAST experiment~\cite{cast}. Manifestly, the PVLAS result contradicts this bound by about five orders of magnitude. A possible way out of
this difficulty is to suppose that, because of some new physics, the LPB detected by PVLAS is trapped inside star cores. A specific scenario in which this indeed occurs has been put forward~\cite{masso2005}. While no fully satisfactory solution to this problem has yet been found, the existence of new physics at low energy is evidently called for, and this poses an exciting challenge for contemporary research. Thus, the need for an independent test of the PVLAS result looks imperative. 

In this Letter, we show that LPBs can give rise to observable
effects if suitably selected binary pulsars are monitored in the
gamma-ray band with future satellite-borne detectors like GLAST. Our
analysis is carried out for arbitrary values of $m$ and $M$, and the
requirement of observability defines the corresponding parameter ranges. It
turns out that the values found by PVLAS fall within these
ranges, and so our result provides a cross-check for the PVLAS
claim of a new LPB. Actually, our calculations apply also to the case
of scalar bosons, except that the ${\bf E} \cdot {\bf B}$ coupling
should be replaced by the ${\bf E} \cdot {\bf E} - {\bf B} \cdot {\bf B}$ coupling.

In what follows we will focus, as a model system,
on the specific case of the recently-discovered binary pulsar system J0737-3039
which is seen almost edge on \cite{DoublePulsar} and for which gravitational
microlensing has also been studied \cite{lai}. We have already shown \cite{dupays} that
in the peculiar geometry of this system it will be possible to observe
quantum vacuum lensing resulting from the photon-magnetic field
interaction, with photon energies in the X-ray domain. There
we used the same system, again starting with the photon beam from one of two
stars (pulsar $A$). Denote by $\rho$ the
impact parameter of the beam, namely its projected distance from the
companion, pulsar $B$. When $\rho$ is small,
the beam goes through the magnetosphere of pulsar $B$. Therefore the observed photons traverse a region where a strong magnetic field is present. Assume now that a LPB exists. Then, owing to 
its two-photon coupling (1), some of the photons will be converted
into LPBs, thereby decreasing the beam intensity. Depending on the
actual values of the physical parameters involved, the beam
attenuation can be substantial. As we will see, the effect turns out
to be important for photons with energy larger that $10 \, {\rm MeV}$,
namely in the gamma-ray band. Remarkably enough, these photons
propagate totally unimpeded in the magnetosphere of pulsar $B$, and so
we do not have to bother about further potential sources of beam
attenuation. Moreover, the effect under consideration does not depend
on any new physics responsible for the confinement of the LPBs in the
star interiors \cite{masso2005}.

\begin{figure}[h]
\includegraphics[width=8cm]{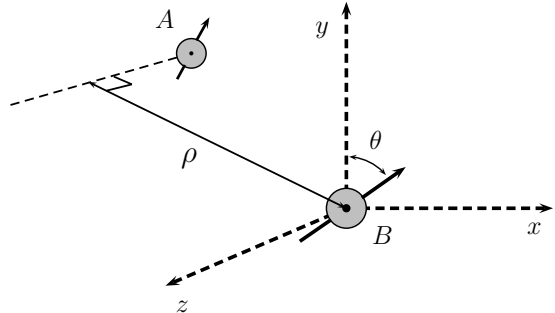}
\caption{\label{fig0}{Geometry of the model binary pulsar system}}
\end{figure}

The model system studied, 
J$0737-3039$, is particularly well suited for our purposes. Its
orbital plane has currently an inclination angle $i \simeq
87^0$~\cite{DoublePulsar}, and the orbital precession of the system
allows us to predict that the value $i \simeq 90^0$ could be achieved
before 2020. The orbital period is 2 hours and 45 minutes, while the
spinning period of pulsar $B$ is 2.77 seconds.  Although the current minimum value of the impact parameter of the light beam from pulsar $A$ is somewhat uncertain, a realistic estimate yields $\rho \simeq 4 \cdot 10^3 \, {\rm km}$~\cite{coles} and this is the value used in our analysis. 
The geometry of the binary pulsar under consideration is depicted in Fig. (\ref{fig0}). 

The light beam from pulsar $A$ propagates along the $z$ direction and
passes pulsar $B$ at impact parameter $\rho$. We introduce a $(x,y,z)$
coordinate system fixed with pulsar $B$. For simplicity, we suppose that the rotation axis of pulsar $B$ lies along the $y$ direction and that its magnetic dipole moment
precedes at an angle $\theta = 45^0$ about the $y$ axis. The light beam from pulsar $A$ experiences a time-dependent and 
inhomogeneous dipolar magnetic field ${\bf B}$. The photon-LPB oscillations are
described by the coupled Klein-Gordon and Maxwell equations dictated by
lagrangian (\ref{eq1}), plus free-field terms. In the short-wavelength approximation (to be justified later), the propagation equations take the following first-order form~\cite{Raffelt and Stodolsky}
\begin{equation}
\label{eq2}
i \, \frac{\partial}{\partial z} | \psi (z) \rangle = {\cal M} | \psi (z) \rangle~,
\end{equation}
with
\begin{equation}
\label{eq3}
| \psi (z) \rangle \equiv c_x (z) \, |x \rangle + c_y (z) \, |y \rangle + c_{\phi} \, 
| \phi \rangle~,        
\end{equation} 
where $| x \rangle$ and $| y \rangle$ are the two linear polarization states of the photon (propagating in the $z$-direction) along the $x$ and $y$ axis, respectively, and $| \phi \rangle$ is the LPB state. Moreover, the mixing matrix ${\cal M}$ is given by
\begin{equation}
\label{eq4}
{\cal M} = \left(
\begin{array}{ccc}
\omega + \Delta_{xx} & \Delta_{xy} & B_x /2M \\
\Delta_{yx} & \omega + \Delta_{yy} & B_y /2M \\
B_x /2M & B_y /2M & \omega - m^2 /2 \omega \\
\end{array}
\right)~,
\end{equation}
with $\omega$ denoting the photon/LPB energy. While the terms appearing in the third row and column of ${\cal M}$ have an evident physical meaning, the $\Delta$-terms require further
discussion. Generally speaking, they reflect both the properties of the medium and the QED vacuum polarization effects. Off-diagonal terms $\Delta_{xy}, \Delta_{yx}$ directly mix the photon polarization states and typically give rise to Faraday rotation. Clearly they are
important when the polarization of photon beam is taken into account,
but become irrelevant for the present analysis. The diagonal elements $\Delta_{xx}, \Delta_{yy}$ arise from two independent contributions: vacuum polarization effects and plasma effects. The first ones are due to one-loop Feynman diagrams and are analytically described by the
Heisenberg-Euler effective lagrangian when terms involving LPBs in the loops are discarded, as is currently done. Neglecting again beam polarization effects, we get ${\Delta}^{QED}_{xx} \simeq {\Delta}^{QED}_{yy} \simeq ({\alpha} /45 \pi ) ( B / B_{cr})^2 \, \omega$, where $\alpha$ is the fine-structure constant and $B_{cr} \simeq 4.41 \cdot 10^{13} \, {\rm G}$ denotes the critical magnetic field~\cite{adler}. Plasma effects arise from the presence of free charges in the magnetosphere of pulsar $B$ and their contribution is ${\Delta}^{PL}_{xx} \simeq 
{\Delta}^{PL}_{yy} \simeq - \, 2 \pi \alpha \, n_e / \omega \, m_e$,
where $m_e$ is the electron mass and $n_e$ is the electron density~\cite{Raffelt and Stodolsky}. Off-diagonal QED effects and plasma effects turn out to be presently irrelevant and will be neglected (more about this, later). 

We solve eq. (\ref{eq2}) numerically in the equivalent exponential form 
\begin{equation}
\label{eq5}
\ket{\psi(z)}={\cal{T}}\exp{\left(-i\int_{z_0}^{z}{{\cal M}(z^\prime)dz^\prime}\right)}\ket{\psi(z_0)}~,
\end{equation} 
where ${\cal{T}}$ is Dyson's ordering operator. Since the $\cal{M}$ matrix is explicitly 
$z$-dependent, the propagation is computed iteratively, i.e. in steps of small intervals $\delta
z$. For small enough $\delta z$, the $\cal{M}$ matrix can be considered as constant over one distance step, and the short-propagation iterations can be calculated as \cite{Heather:91}
\begin{equation}
\label{eq6}
\ket{\psi(z+\delta z)}\simeq\exp{\left(-i{\cal M}(z)\delta z\right)}\ket{\psi(z)}~.
\end{equation}

The propagation is performed from $z_0$ to $z_{max}$ with the initial condition corresponding 
to a pure photon state. Here $z_0$ is fixed by the trajectory of pulsar $A$ and $z_{max}$ is chosen so that $B_x(z) /2M \simeq B_y(z) /2M \simeq 0$ for $z>z_{max}$. The transition probability is then 
\begin{equation}
\label{eq7}
P_{(\gamma_j \to \phi)} = |\brk{\phi}{\psi(z_{max})}|^2 = |c_\phi(z_{max})|^2~,
\end{equation}
where $j=x$ (resp. $y$) refers to the initial photon polarization along the $x$-axis
(resp. $y$-axis). For an unpolarized photon beam, the transition probability is  
\begin{equation}
\label{eq8}
P_{(\gamma \to \phi)} = \frac{P_{(\gamma_x \to \phi)}}{2}+\frac{P_{(\gamma_y \to \phi)}}{2}~.
\end{equation}

In what follows we use the values of $m$ and $M$ as suggested by the
PVLAS result. In Fig. 2, we have plotted the transition probability $P_{(\gamma \to \phi)}$ versus the photon energy $\omega$, for a gamma beam with impact parameter $\rho = 4 \cdot 10^3 \, {\rm km}$.

\begin{figure}[h]
\includegraphics[width=8cm]{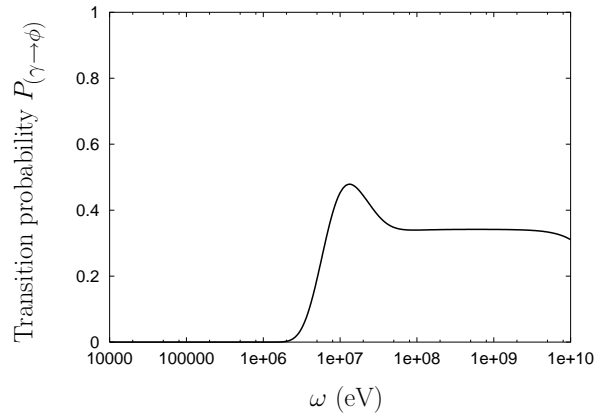}
\caption{\label{fig1}{Transition probability versus photon energy for a trajectory of the light beam with an impact parameter $\rho=4 \cdot 10^3$ km.}}
\end{figure}

The strong energy-dependence exhibited by $P_{(\gamma \to \phi)}$ suggests to investigate the photon-LPB conversion in the region of the broad maximum, namely for $\omega > 10 \, {\rm MeV}$. Observe that this circumstance fully justifies the above short-wavelength approximation.

Actually, the threshold behaviour of $P_{(\gamma \to \phi)}$ can be
understood as follows. Consider a drastically simplified situation in which the magnetic field experienced by the light beam is
stationary and homogeneous. A typical value of the magnetic field strenght on the surface of a neutron star, namely at a distance $r \simeq 10 \, {\rm km}$ from the centre, is $B_0 \simeq 
10^{12} \, {\rm  G}$. Since ${\bf B}$ is dipolar, it goes like $\sim r^{- 3}$, and so
the magnetic field strenght at $\rho \simeq 4 \cdot 10^{3} \, {\rm km}$ is $B \simeq 1.6 \cdot 
10^{4} \, {\rm G}$. Correspondingly, we have ${\Delta}^{QED}_{xx} \simeq {\Delta}^{QED}_{yy} \simeq 6.3 \cdot 10^{  - 18} \, (\omega /{\rm MeV}) \, {\rm eV}$. Because the terms
$\Delta_{xy}, \Delta_{yx}$ presently get only a QED contribution, they should not exceed 
${\Delta}^{QED}_{xx} \simeq {\Delta}^{QED}_{yy}$. Following ref. \cite{Raffelt and Stodolsky}, it
is straightforward to see that the terms ${\Delta}^{PL}_{xx} \simeq
{\Delta}^{PL}_{yy}$ are much smaller than ${\Delta}^{QED}_{xx} \simeq
{\Delta}^{QED}_{yy}$ for $B \simeq 1.6 \cdot 10^{4} \, {\rm G}$. We find $B_x / 2 M \simeq B_y
/2 M \simeq 0.4 \cdot 10^{- 12} \, {\rm eV}$ and $m^2 / 2 \omega
\simeq 0.5 \cdot 10^{- 12} \, ({\rm MeV}/ \omega) \, {\rm eV}$. Correspondingly $\Delta_{xy}$ and $\Delta_{yx}$ turn out to be the smallest elements of ${\cal M}$. Thus, we see that off-diagonal QED effects as well as plasma effects are indeed negligible (in agreement with previous assumptions). In addition, mixing effects in ${\cal M}$ should become physically important when the ratio of the off-diagonal elements to
the difference of the diagonal ones starts to be of order 1. Therefore, photon-LPB conversion is expected to actually occur for $\omega > 1 \, {\rm MeV}$, which means that a substantial beam
attenuation ought to show up when gamma-ray photons above $10 \, {\rm MeV}$ are observed.

\begin{figure}[h]
\includegraphics[width=8cm]{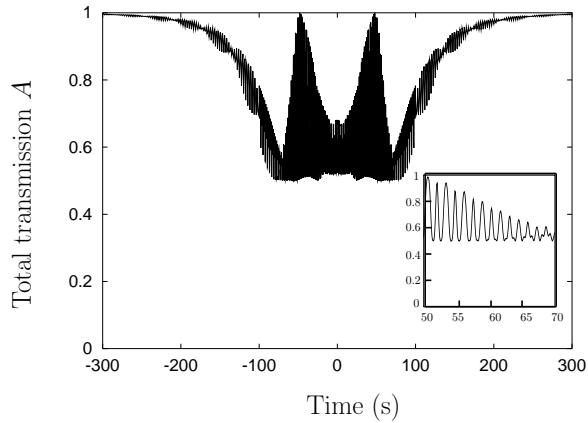}
\caption{\label{fig2}{Total transmisson of the gamma photon beam emitted by
    pulsar $A$ versus
    time. Inset shows the modulation mainly due to the rotation of the
    magnetic dipole moment of pulsar $B$.}}
\end{figure}

It makes even more sense, at this point, to consider the case of
J0737, because pulsars are frequently celestial gamma-ray sources \cite{Bignami:96}, because of the very special geometry of this system and because the NASA GLAST mission (due to launch in 2007) will have a pulsar gamma-ray detection sensitivity up to two orders of magnitude better than any previous telescope \cite{Thomson}.

Assuming that the gamma photon beam emitted by pulsar $A$ is not polarized, the temporal behaviour of the considered effect is best expressed in terms of the total {\it transmission} $A = 1 - P_{(\gamma \to \phi)}$ of the gamma ray after propagation in the magnetosphere of pulsar $B$. In Fig. 3, we plot $A$ versus time as pulsar $A$ moves in its nearly edge-on orbit. Our numerical simulation predicts in this case a strong attenuation of the photon beam up to
$50\,\%$ with a time duration of about $200$ s. The effect has three
different temporal structures. Of course, the broad minimum, from
$-200$ s to $+200$ s, corresponds to the transit of pulsar $B$ "in
front" of pulsar $A$. The tens-of-seconds, symmetric peaks are due to photon-LPB 
oscillations depending on the path
through the interaction region with pulsar $B$. Finally, the highest
frequency modulation, shown
in inset, is due to the rotation of the magnetic dipole moment of
pulsar $B$. Under the assumptions corresponding to the PVLAS result, a macroscopic effect could be observable from the
J0737-3039 system, given enough instrumental sensibility and observing
time. We note that pulsar $A$ also generates a similar effect on the
gamma photon beam emitted by pulsar $B$. Since the spin period of pulsar $A$ is smaller than the one of pulsar $B$ ($23$ ms), the modulation period of the effect will be also smaller.

\begin{figure}[h]
\includegraphics[width=8cm]{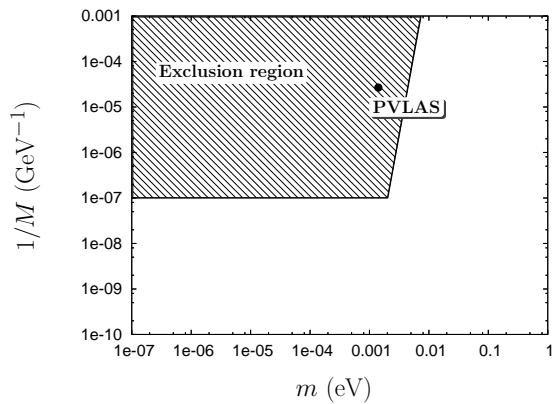}
\caption{\label{fig3}{Exclusion region in the case that the existence of the attenuation is excluded at $10\,\%$ level.}}
\end{figure}

Finally, Fig. 4 shows the region in the $m-M$ parameter space which is excluded by no detection of an attenuation $A$ at the $10 \, \%$ level.
To get such an attenuation, one need to count 100 photons during the
total integrating time. For two weeks observation time, this
corresponds to a flux from pulsar $A$ of about $2 \cdot 10^{-7}$ photons/cm$^2$/s which is a reasonable flux according to previous observations
of several pulsars \cite{Bignami:96}. We note in any case that the
GLAST sensibility curves allow for a much weaker minimum detectable
flux, down to $1 \cdot 10^{-10}$ photons/cm$^2$/s \cite{Glastlat}.
In conclusion, we have shown that LPBs can produce observable effects
in a binary pulsar system seen almost edge-on, like J0737-3039. The
effect is in the gamma-ray band, depending of course on the actual
values of the parameters $m$ and $M$. This is an intriguing
feature of our result, since
pulsars are frequently celestial gamma-ray sources and the NASA GLAST
mission (due to launch in 2007) will have a pulsar gamma-ray detection
sensitivity up to two orders of magnitude better than any previous
telescope \cite{Thomson}. Of course, new binary pulsar systems will be discovered soon through radio-astronomy or possibly by the very GLAST mission. A potentially useful new tool seems to be coming of age.

\end{document}